\documentclass[conference]{IEEEtran}
\IEEEoverridecommandlockouts
\usepackage{cite}
\usepackage{amsmath,amssymb,amsfonts}
\usepackage{algorithm2e}
\usepackage{graphicx}
\usepackage{textcomp}
\usepackage{xcolor}
\def\BibTeX{{\rm B\kern-.05em{\sc i\kern-.025em b}\kern-.08em
    T\kern-.1667em\lower.7ex\hbox{E}\kern-.125emX}}
\usepackage[binary-units]{siunitx}
\DeclareSIUnit{\bps}{bps}
\usepackage[caption=false,font=footnotesize]{subfig}
\usepackage{tabularx}
\usepackage{multirow}

\begin{document}
\title{Initial Beamforming for Millimeter-Wave and Terahertz Communications in 6G Mobile Systems}

\author{
	\IEEEauthorblockN{Wei~Jiang\IEEEauthorrefmark{1}\IEEEauthorrefmark{2},  and Hans~D.~Schotten\IEEEauthorrefmark{2}\IEEEauthorrefmark{1}}
	\IEEEauthorblockA{\IEEEauthorrefmark{1}Intelligent Networking Research Group, German Research Centre for Artificial Intelligence (DFKI), Germany\\
		\IEEEauthorrefmark{2}Institute for Wireless Communication and Navigation, Technische Universit\"at (TU) Kaiserslautern, Germany
	}%
}
\maketitle

\begin{abstract}
 To meet the demand of supreme data rates in terabits-per-second, the next-generation mobile system needs to exploit the abundant spectrum in the millimeter-wave and terahertz bands. However, high-frequency transmission heavily relies on large-scale antenna arrays to reap high beamforming gain, used to compensate for severe propagation loss. It raises a problem of omni-directional beamforming during the phase of initial access, where a base station is required to broadcast synchronization signals and system information to all users within its coverage. This paper proposes a novel initial beamforming scheme, which provides instantaneous gain equally in all directions by forming a pair of complementary beams.  Numerical results verify that it can achieve omni-directional coverage  with the optimal performance that remarkably outperforms the previous scheme called random beamforming. It is applicable for any form of large-scale arrays, and all three architecture, i.e., digital, analog, and hybrid beamforming.
\end{abstract}

\section{Introduction}
Although the fifth-generation (5G) mobile technology is still on its way being deployed across the world, academia and industry have already shifted their focus towards the next-generation technology known as the sixth generation (6G).  Major mobile-industry players have initiated their activities to identify 6G use cases and requirements  \cite{Ref_jiang2022kickoff}. For example, the ITU-T established a focus group \textit{Technologies for Network 2030} in July 2018, envisioning that 6G will support disruptive applications, such as holographic communications, extended reality, Tactile Internet, artificial intelligence \cite{Ref_jiang2017experimental}, and digital twin.  These impose extreme capacities and performance requirements, including a peak data rate of 1 terabits-per-second (Tbps), a massive connection density of $10^7$ devices per \si{\kilo\meter^2}, and an area traffic capacity of \SI{1}{\giga\bps\per\meter^2}, as envisioned by the wireless research community  \cite{Ref_jiang2021road}.

Although the millimeter-wave (mmWave) band has been employed in 5G, the available spectral resources are still very limited relative to the demand of 6G. The ITU-R has assigned a total of \SI{13.5}{\giga\hertz} spectrum for mmWave communications in World Radiocommunication Conference 2019 (WRC-19). As a follow-up, 3GPP specified the Second Frequency Range (FR2) of 5G New Radio (NR), covering \SI{24.25}{\giga\hertz} to \SI{52.6}{\giga\hertz}. Using a bandwidth on the order of magnitude around \SI{10}{\giga\hertz}, a data rate of \SI{1}{\tera\bps} can only be achieved under a spectral efficiency approaching \SI{100}{\bps/\hertz}, which requires symbol fidelity that is not feasible using currently known digital modulation techniques or transceiver components. Consequently, 6G has to exploit the massively abundant spectrum at the high-frequency band above \SI{100}{\giga\hertz}. This band refers to the high end of mmWave, which is defined as \SIrange{30}{300}{\giga\hertz}, or the low end of terahertz (THz) usually covering \SIrange{0.1}{3}{\tera\hertz}. At the WRC-19, the ITU-R has already identified the open of the spectrum between \SI{275}{\giga\hertz} and \SI{450}{\giga\hertz} to land mobile and fixed services, paving the way of deploying THz commutations in 6G \cite{Ref_kuerner2020impact}.

Despite its high potential, mmWave and THz communications suffer from severe propagation losses raised from high free-space path loss, atmospheric gaseous absorption, rainfall attenuation, and blockage \cite{Ref_siles2015atmospheric}, leading to a very short transmission range. Therefore, highly directional antennas are required at the base station and/or the mobile terminal to achieve sufficient beamforming gains to compensate for such losses. Nevertheless, a beam-based system encounters the problem of \textit{initial access} \cite{Ref_giordani2016initial}. In any cellular system, when a terminal powers on, or performs the transition from the IDEL to CONNECTED mode, it needs to search for a suitable cell to access. Meanwhile, a terminal needs to detect the neighboring cells of its serving cell to prepare for handover. To this end, base stations must periodically broadcast synchronization signals and system information in the downlink with \textit{omnidirectional coverage}. However, mmWave and THz systems rely on pencil beams to transmit both control and user data, whereas omni-directional beamforming covering mobile users at any direction is still an open issue \cite{Ref_barati2016initial}.

State-of-the-art methods for omni-directional coverage and the standardized initial-access procedures in 4G LTE and 5G NR are given in the next section. Among all existing methods, the most appropriate technique for initial access in mmWave and THz communications is called \textit{random beamforming} (RBF) proposed by Yang and Jiang in \cite{Ref_yang2013random, Ref_yang2012methodUS8170132, Ref_jiang2012methodUS13685426, Ref_yang2013methodUS8537785, Ref_jiang2012enhanced, Ref_yang2012methodUS13654743}. But it still has some drawbacks. Hence, this paper proposed a novel initial-access method called \textit{complementary beamforming} (CBF), which goes beyond the RBF in terms of the following aspects:
\begin{itemize}
    \item The RBF achieves omni-directional coverage by \textit{averaging} many beams over the time or frequency domain. There is a performance gap compared to the benchmark (the single-antenna broadcasting) due to the energy fluctuation in the angular domain. In contrast, the CBF can provide \textit{instantaneously} equal gain using a pair of complementary beams and thus improve performance.
    \item The RBF needs to form sufficient random beam patterns in a short time/frequency span to achieve averaged equal gains over all directions. The isotropic gain of the CBF is achieved by a pair of complementary beams, which can be fixed. Thus, it simplifies the implementation.
    \item The previous work \cite{Ref_yang2013random} discussed beamforming over a uniform linear array (ULA) with only a few elements. However, mmWave and THz communications need to utilize large-scale antenna arrays with a massive number of elements to achieve high power gain. Therefore, this paper expands the discussion to large-scale arrays such as a uniform planar array (UPA).
    \item The discussion of \cite{Ref_yang2013random} focused on \textit{digital beamforming}, but it is too expensive and power-angry for large-scale arrays. The CBF adapts to all three types architecture, i.e., digital beamforming, analog beamforming, and hybrid beamforming.
\end{itemize}

The remainder of this paper is organized as follows: Section II provides an overview of related technologies, including the initial access in 4G LTE and 5G NR. Section III introduces the system model for beamforming, while Section IV explains the proposed complementary beamforming. Finally, Section V gives some examples of numerical results, and conclusions are made in Section VI.

\section{State-of-the-art Initial Access Technologies}
The traditional approach of transmitting synchronization and broadcast signals is to use a single antenna that has an omnidirectional radiation pattern. Consequently, signal broadcasting was never a concern in earlier mobile systems that employed single-antenna base stations. However, an antenna array has become an essential part of advanced wireless systems to improve spectral efficiency. Intuitively, one can select a specific antenna in an array to transmit the broadcast signals. Nevertheless, the selected antenna needs a much higher power amplifier (PA) \cite{Ref_jiang2012suppressing}, which is more expensive and power-consuming than the remaining antennas to achieve similar coverage as the unicast signal leveraging a beamforming gain. Therefore, it is meaningful to re-use multiple low-powered antennas to transmit broadcast signals to guarantee a cost-efficient and power-efficient system \cite{Ref_yang2013random}.

There have been several such multi-antenna schemes for omnidirectional coverage. A space-time block code, particularly the Alamouti code, has been successfully applied in the Universal Mobile Telecommunications System (UMTS) in the case of two transmit antennas.  In the TD-SCDMA system, where the smart-antenna technique was used, a particular beam, known as \textit{the broadcast beam}, with a flat amplitude within a certain angular range, such as \SI{120}{\degree}, was designed for broadcast channels. However, this scheme has two deficiencies. The first is the low power efficiency caused by small weight coefficients. For example, a broadcast beam with \SI{120}{\degree} is generated by the weighting vector $[0.55, 1, 1, 0.55, 0.85, 1, 0.85]$, where two Radio-Frequency (RF) channels have utilized approximately only $30\%$ (due to $0.552^2=0.304704$) of the full capability of the PA. The second is the adverse deformation of the broadcast beam stemmed from the character deviation and failure of RF channels. Cyclic delay diversity (CDD) is a simple multi-antenna transmission scheme recommended for Digital Video Broadcasting (DVB) and 3GPP LTE. Some research studies revealed that CDD is essentially a beamforming technique in the frequency domain. However, the performance of CDD has not yet been theoretically proved and generally verified by practices, particularly in cases of four antennas and above.

\begin{figure}[!htbp]
\centering
    \includegraphics[width=0.4\textwidth]{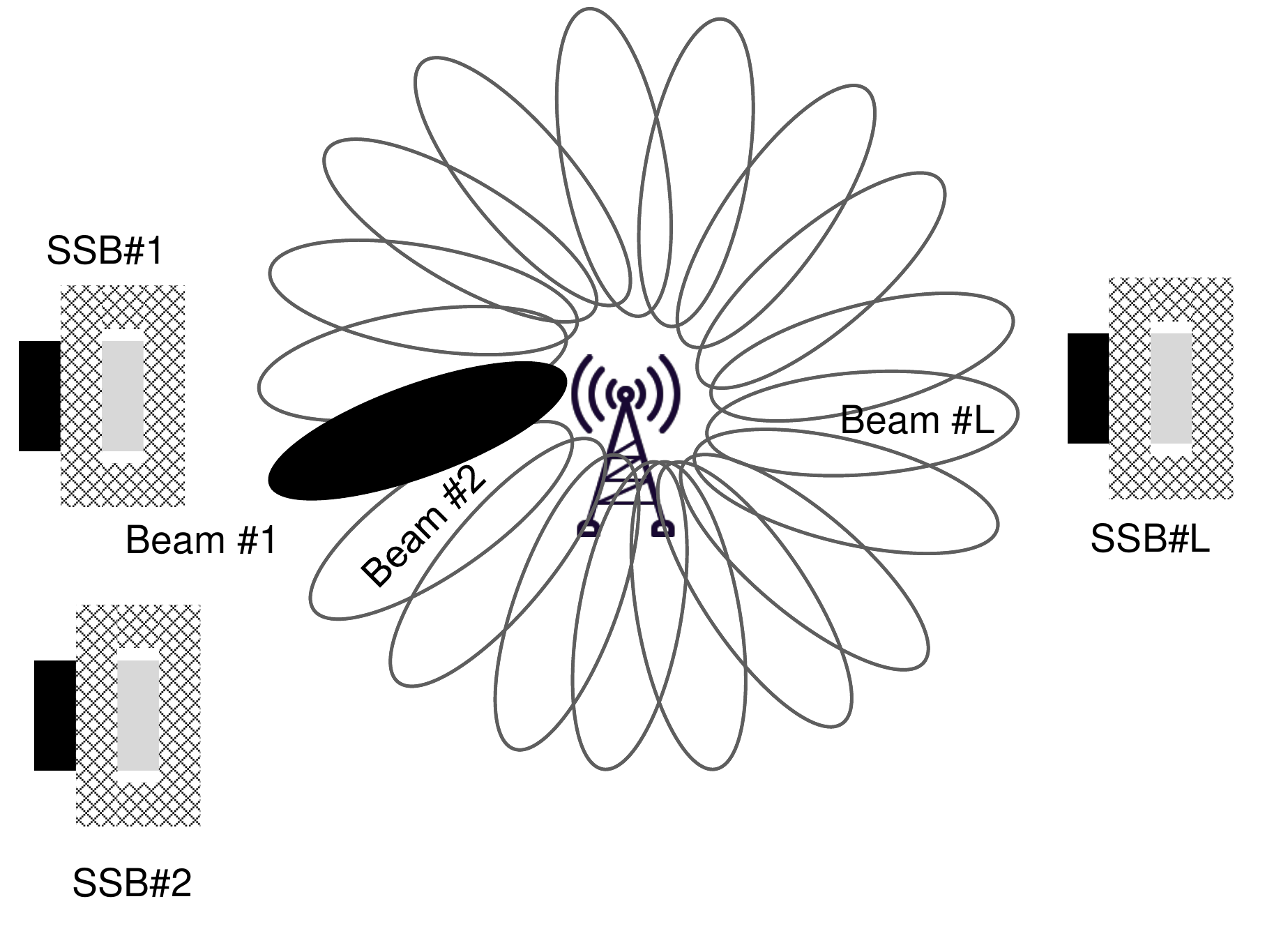}
    \caption{Omni-directional beam sweeping adopted for initial access in 5G NR, where each beam transmits the same SSB and up to $L=64$ beams are supported in the FR2 band to cover the \SI{360}{\degree} angular space \cite{Ref_dahlman20215gNR}. }
    \label{fig:NR:SSB}
\end{figure}

5G NR first defined the term Synchronization Signal Block (SSB), consisting of the Primary Synchronization Signal (PSS), Secondary Synchronization Signal (SSS), and Physical Broadcast Channel (PBCH). The initial access of an NR system is achieved by a brute-force method called beam sweeping to sequentially scan the \SI{360}{\degree} angular space with multiple narrow beams.
Both terminals (UE) and base stations (gNodeB) have a predefined codebook consisting of a set of weighting vectors, where each weighting vector can form a narrow beam to cover a particular direction, and all beams together seamlessly cover the whole angular space, as demonstrated in \figurename \ref{fig:NR:SSB}. The set of SSBs within a beam-sweeping procedure to scan \SI{360}{\degree} is referred to as an \textit{SS burst set}. An NR system operating in the FR2 band can support up to $L=64$ SSBs within a burst set using \num{64} beams for sweeping. An SSB is transmitted over each beam to guarantee that all directions can receive synchronization signals and master system information.  This method provides good coverage but suffers from high overhead (the SSB transmission repeats $L$ times) and long discovery delay.

\section{System Model}
The previous work \cite{Ref_yang2013random} uses a ULA to discuss the principle of random beamforming. Assuming that there are $N$ elements with inter-element spacing of $d$, the steering vector is
\begin{equation}
    \textbf{a}(\theta) = \left[1, e^{-j \frac{2\pi}{\lambda} d\sin \theta}, \ldots,e^{-j \frac{2\pi}{\lambda} (N-1)d\sin \theta} \right]^T.
\end{equation}
Using a weighting vector on time-frequency unit $t$
\begin{equation}
    \textbf{w}(t) = \left[w_1(t),w_2(t),\ldots,w_N(t) \right]^T.
\end{equation}
Then, the beam pattern of ULA can be given by
\begin{equation}
    g(\theta,t) = \textbf{w}^H(t)\textbf{a}(\theta).
\end{equation}

\begin{figure}[!bpht]
\centering
\includegraphics[width=0.35\textwidth]{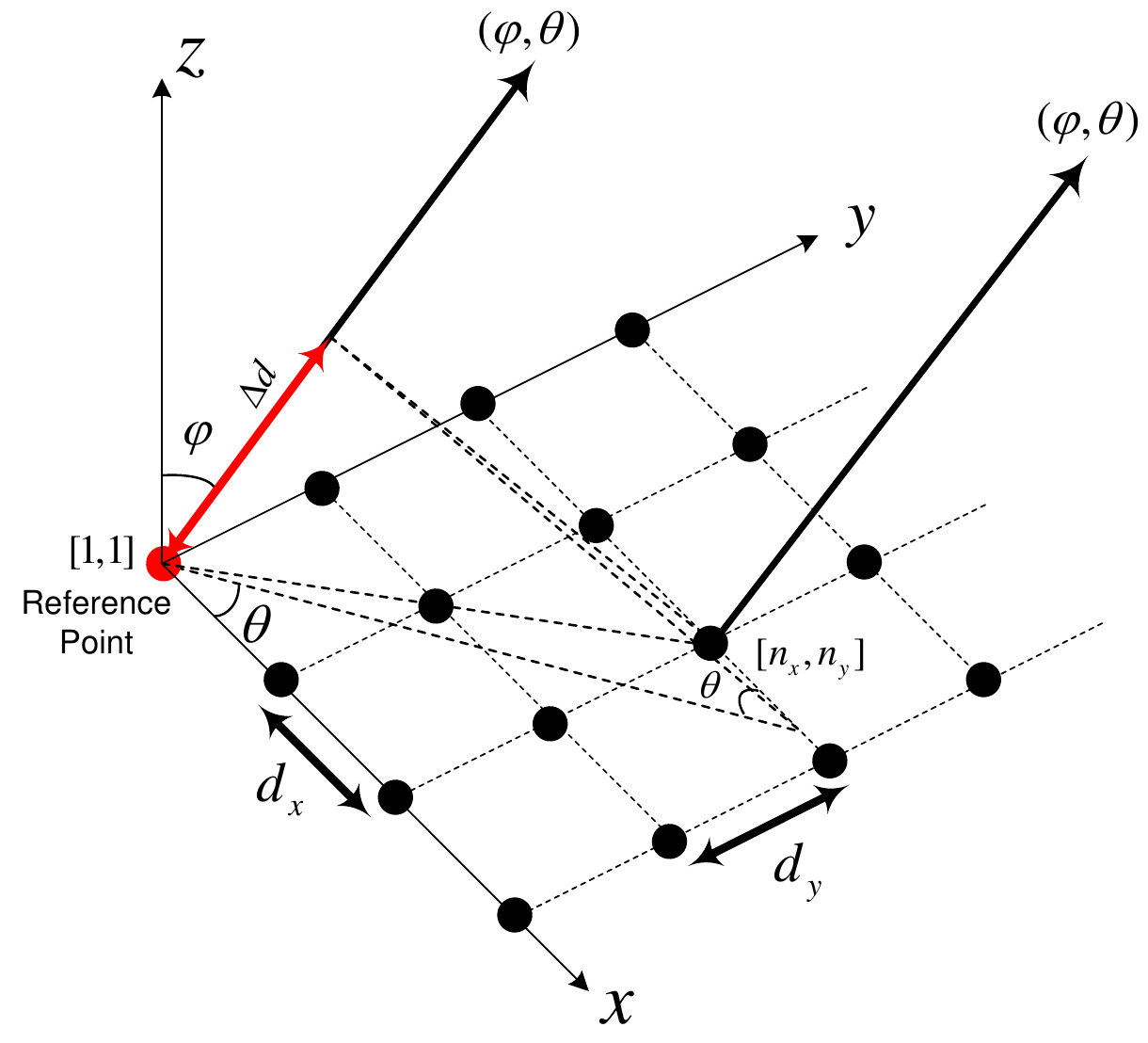}
\caption{Far-field geometry of the radiation of a plane wave from a UPA in the direction with an elevation angle $\varphi$ and an azimuthal angle $\theta$. }
\label{Diagram_planararray}
\end{figure}

However, an array used for high-frequency transmission usually has a large number of elements, e.g., \numrange{64}{256} antennas at the base-station side specified by 5G NR. One of the major challenges is how to pack these elements within a restricted volume. The solution is to use planar arrays such as a uniform planar array. As illustrated in \figurename \ref{Diagram_planararray}, a UPA having $N_x$ rows in the $x$-axis with inter-element spacing of $d_x$, and $N_y$ columns in the $y$-axis separated by $d_y$. In a three-dimensional (3D) coordinate system, the direction of arrival or departure of a plane wave is described by an elevation angle $\varphi$ and an azimuthal angle $\theta$, denoted by $(\varphi, \theta)$.

The difference of propagation distance between the reference point $[1,1]$ and a typical element located in the $n_x^{th}$ row and the $n_y^{th}$ column is given by
\begin{equation} \nonumber
    \triangle d_{n_xn_y}(\varphi,\theta) = (n_x-1)d_x \sin \varphi \cos\theta + (n_y-1)d_y \sin \varphi \sin\theta,
\end{equation}
which raises a time difference of
\begin{align}
    \triangle \tau_{n_xn_y}(\varphi,\theta) &=  \frac{\triangle d_{n_xn_y}(\varphi,\theta)}{c}\\ \nonumber
    &= \frac{\sin \varphi \left[ (n_x-1)d_x  \cos\theta + (n_y-1)d_y  \sin\theta \right]}{c},
\end{align}
where $c$ stands for the speed of light.
The 3D steering vector of the UPA is expressed by \cite{Ref_chen2013when}
\begin{equation}
    \textbf{a}(\varphi, \theta) = \textbf{v}_x(\varphi, \theta) \otimes \textbf{v}_y(\varphi, \theta),
\end{equation}
where $\textbf{v}_x(\varphi, \theta)$ and $\textbf{v}_y(\varphi, \theta)$ can be viewed as the steering vectors on the x- and y-direction, respectively, with
\begin{equation} \nonumber
    \textbf{v}_x(\varphi, \theta) = \left[1, e^{j\frac{2\pi}{\lambda} d_x \sin \varphi \cos \theta }, \ldots,e^{j\frac{2\pi}{\lambda} (N_x-1)d_x \sin \varphi \cos \theta } \right]^T,
\end{equation}
and
\begin{equation} \nonumber
    \textbf{v}_y(\varphi, \theta) = \left[1, e^{j\frac{2\pi}{\lambda} d_y \sin \varphi \sin \theta }, \ldots,e^{j\frac{2\pi}{\lambda} (N_y-1)d_y \sin \varphi \sin \theta } \right]^T.
\end{equation}

We write $w_{n_xn_y}(t)$ to denote the weighting coefficient for element $[n_x,n_y]$ and assume the planar array transmits a single-stream narrow-band signal $s(t)$ to a typical user. The receiver observes an incoming signal as given in (\ref{eqn:mmWave:RXsignalPlanaryArray}). Then, we can get the 3D beam pattern as expressed by (\ref{Eqn_UPAbeam}) at the top of the next page.
\begin{figure*}[!t]
\begin{align} \nonumber \label{eqn:mmWave:RXsignalPlanaryArray}
    y(t) & = \sum_{n_x=1}^{N_x} \sum_{n_y=1}^{N_y} w_{n_xn_y}(t) s(t)e^{j2\pi f_0\left[t-\tau_{n_xn_y}(\varphi,\theta)\right]} + n(t)\\
         &= \left(\sum_{n_x=1}^{N_x} \sum_{n_y=1}^{N_y} w_{n_xn_y}(t) e^{-j\frac{2\pi\sin \varphi}{\lambda}  \left[ (n_x-1)d_x  \cos\theta + (n_y-1)d_y  \sin\theta \right]  } \right) s(t) e^{j2\pi  f_0t} + n(t) = g(\varphi,\theta,t)s(t) e^{j2\pi  f_0t} +n(t)
\end{align} \rule{\textwidth}{0.1mm}
\begin{align} \label{Eqn_UPAbeam}
    g(\varphi,\theta,t)
         &= \sum_{n_x=1}^{N_x} \sum_{n_y=1}^{N_y} w_{n_xn_y}(t) e^{-j\frac{2\pi\sin \varphi}{\lambda}  \left[ (n_x-1)d_x  \cos\theta + (n_y-1)d_y  \sin\theta \right]  }
\end{align} \rule{\textwidth}{0.1mm}
\end{figure*}

\section{Complementary Beamforming}
\subsection{Design Criteria}
The criteria in designing the random pattern sequence in the RBF \cite{Ref_yang2011random} can be summarized as follows
\begin{itemize}
    \item Keep equal average power in each direction for omnidirectional coverage
    \item Set equal power in each antenna to maximize power amplifier efficiency
    \item Use random beams with the minimum variance to achieve the maximal capacity
\end{itemize}
To achieve the objective of equal average power in each direction, \textit{the pattern variance} in the angular dimension is defined as a metric to measure the pattern's degree of deviation from a circle:
\begin{equation} \label{eqn:beamVariation}
    \sigma_g^2 = \int_0^{2\pi}\int_0^{2\pi} \left[ |g(\varphi, \theta,t)|^2 - \mathbb{E}\left( |g(\varphi, \theta,t)|^2\right) \right]^2 d\theta d\varphi .
\end{equation}
To satisfy the second and third criteria, the basis weighting vector is the vector with the minimum variance and unit module for each entry, i.e.,
\begin{itemize}
    \item $|w_1|=|w_2|=\dots=|w_N|=1$,
    \item $\textbf{w}_0=\arg\min (\sigma_g^2)$.
\end{itemize}

The time-frequency resources are divided into many small time-frequency blocks (TFB), and a random pattern is applied on each TFB. For a sufficient number of random patterns, the average power is nearly equal for each direction.
However, the RBF achieves omni-directional coverage with \textit{averaged} equal gain at any direction. The energy distribution of an individual beam still fluctuates for different directions, as \figurename \ref{fig:beam1}. Compared with the single-antenna broadcasting with even energy distribution in the angular domain, this fluctuation leads to performance degradation. To be specific, a beam null such as \SI{0}{\degree} and \SI{180}{\degree} in the figure brings a low signal-to-noise ratio, analog to a deep fade in a wireless channel.

\subsection{Complementary Beams}
For some types of arrays such as a ULA, it is impossible to form a beam with even energy distribution.
To fill this performance gap, this paper proposes to use a pair of beams (or more), rather than a single beam, to remove this constraint. These two beams are complementary each other to form omni-directional coverage with \textit{instantaneous} equal gain at any direction. As a result, the optimal performance identical to that of a single antenna with a high PA can be obtained.

Instead of only a basis weighting vector, the CBF determines a pair of weighting vectors $\textbf{w}_1$ and $\textbf{w}_2$, satisfying the following criteria:
\begin{itemize}
     \item Minimize the variance of individual pattern, i.e.,
        \begin{equation} \nonumber \hat{\textbf{w}}_k=\arg\min(\sigma^2_{g_k}),\: k=1,2, \end{equation}
        where $g_k$ denotes the individual pattern of $\textbf{w}_1$ and $\textbf{w}_2$.
     \item Minimize the variance of the composite pattern, i.e.,
        \begin{equation} \nonumber [\hat{\textbf{w}}_1,\hat{\textbf{w}}_2]=\arg\min(\sigma^2_{g}). \end{equation}
        We define the amplitude of the composite pattern as  \begin{equation}
        |g(\varphi, \theta,t)| =\sqrt{\frac{|g_1(\varphi, \theta,t)|^2+|g_2(\varphi, \theta,t)|^2}{2}}.
    \end{equation}
    \item Set equal transmit power in each antenna to maximize the PA efficiency, i.e., \begin{equation}  |w_1|=|w_2|=\dots=|w_N|=1.\end{equation}
\end{itemize}

\begin{figure*}[!tbph]
\centerline{\hspace{-15mm}
\subfloat[]{\includegraphics[width=0.4\textwidth]{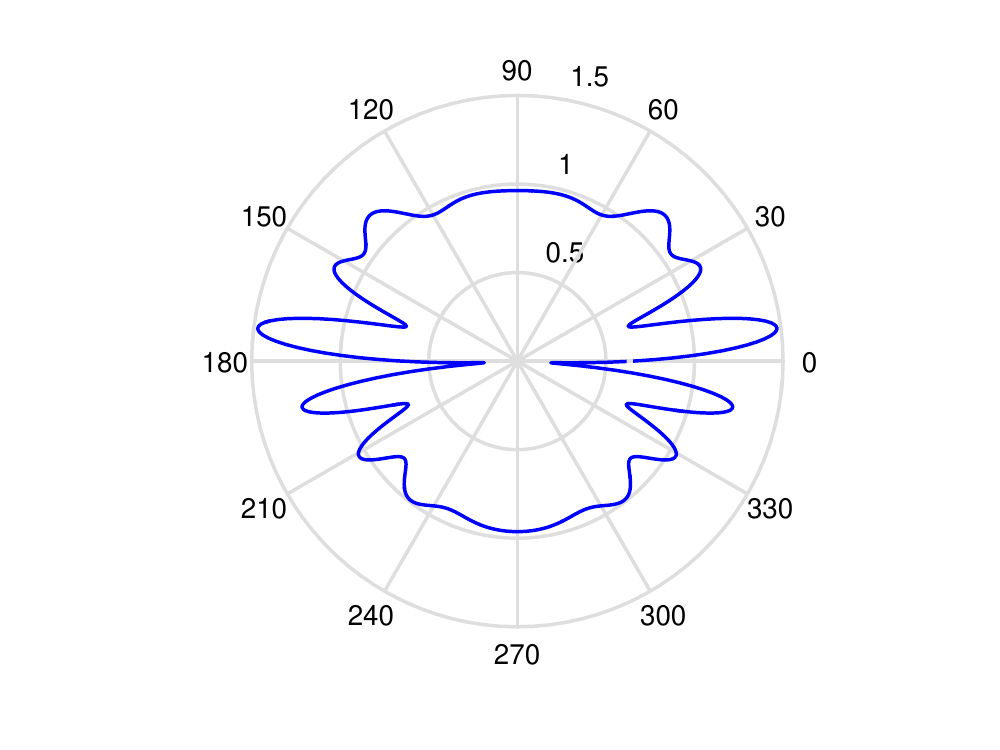}\label{fig:beam1}} \hspace{-14mm}
\subfloat[]{\includegraphics[width=0.4\textwidth]{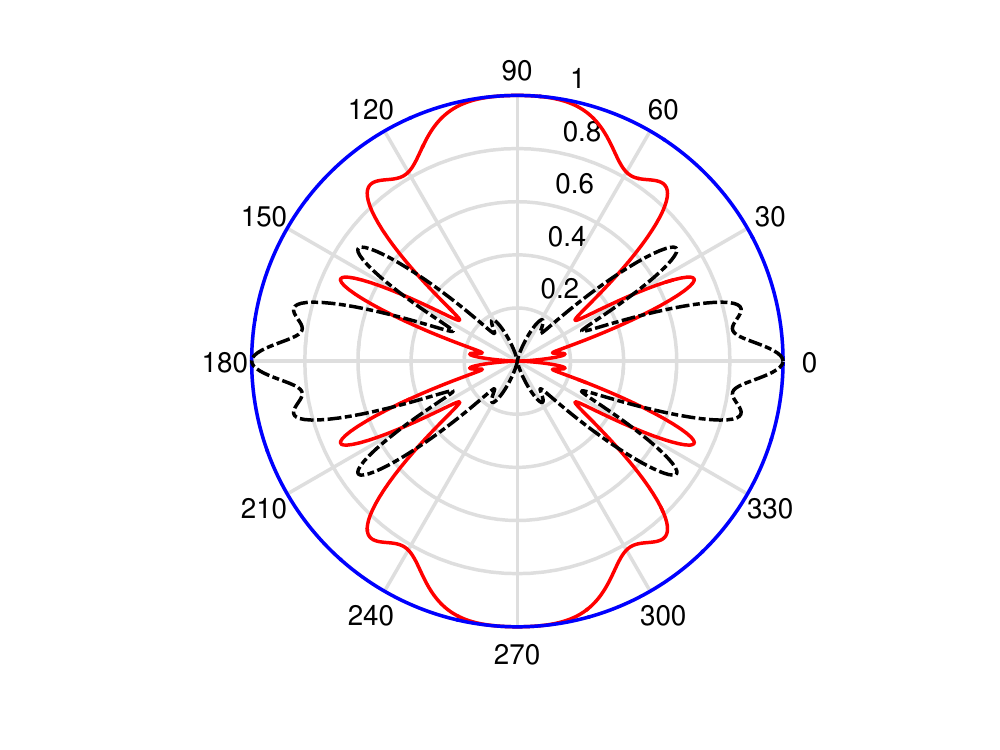}\label{fig:beam2}} \hspace{-14mm}
\subfloat[]{\includegraphics[width=0.4\textwidth]{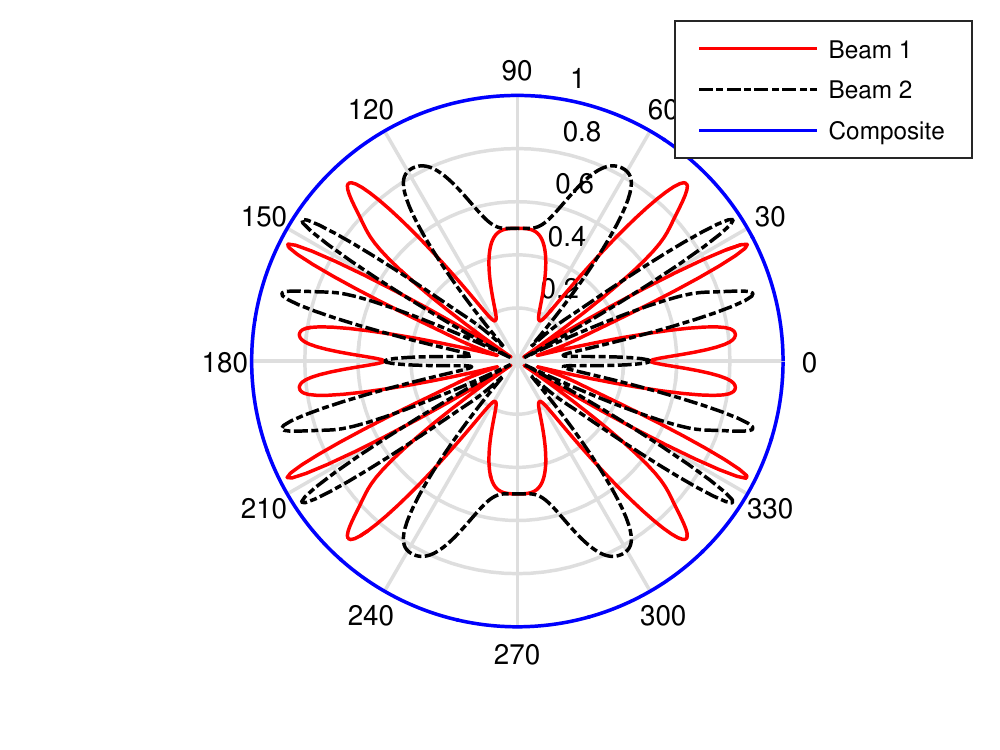}\label{fig:beam3}} }
\hspace{0mm}
 \caption{Comparison of the RBF and CBF: (a) Basis pattern of the RBF using $\textbf{w}=\frac{\sqrt{2}}{2} \left[ -\sqrt{2},-1+i, \sqrt{2}i,1-i,-1+i,1-i,\sqrt{2},1+i \right]^T$ over an 8-element ULA, where $\sigma_g^2=0.3352$; (b) A pair of complementary patterns corresponding to $\textbf{w}_1=[1,-1,   -1,     1,     1,     1,     1,     1]^T$ and $\textbf{w}_2=[-1,     1,    -1,     1,    -1,    -1,     1,     1]^T$, respectively, and their composite pattern is strictly a unit circle ($\sigma_g^2=0$), achieving isotropic instantaneous power as a single antenna with high PA. (c) A pair of complementary patterns over a 16-element ULA, generating a composite pattern with instantaneously equal gain over all directions ($\sigma_g^2=0$).  }
\label{Diagram_BeamPair}
\end{figure*}

Given the steering vector of an array, the weighting vectors for a pair of complementary patterns can be determined by conducting a computer search, as depicted in \algorithmcfname~\ref{alg:001}.
This process is not computationally complex, even with large antenna numbers. Moreover, engineers can figure out these weighting vectors during the process of system design and configure the equipment beforehand, which does not bring any burden on the system operation. For a direct comparison with the RBF \cite{Ref_yang2013random}, we give a pair of complementary beams over an 8-element ULA, as shown in \figurename \ref{fig:beam2}. Although either beam still fluctuates in the angular domain, their composite pattern is strictly isotropic with zero deviation $\sigma_g^2=0$ (as indicated by a unit circle in the figure), namely
\begin{equation}
     |g|^2=\frac{|g_1|^2+|g_2|^2}{2}=1.
\end{equation}

\SetKwComment{Comment}{/* }{ */}
\RestyleAlgo{ruled}
\begin{algorithm}
\caption{Search Complementary Beams} \label{alg:001}
\SetKwInOut{Input}{input}\SetKwInOut{Output}{output} \SetKwInput{kwInit}{Initialization}
\Input{Steering vector $\textbf{a}(\varphi, \theta)$}
\Input{Array dimension $N$}
\Input{Accuracy factor $K>1$}
\KwResult{Complementary weighting vectors $\tilde{\mathbf{w}}_1$,$\tilde{\mathbf{w}}_2$}
Coefficient granularity $\vartriangle\phi \gets \frac{2\pi}{K}$\;
A coefficient set $ \mathbf{C} \gets \{1,e^{\vartriangle \phi},e^{2\vartriangle \phi},\ldots,e^{(K-1)\vartriangle \phi} \}$\;
Space of weighting vectors $\mathbf{W} \gets \mathbf{C}^N$\;
\kwInit { $\tilde{\sigma}^2_g \gets 1$, $\tilde{\mathbf{w}}_1\gets \mathbf{0}$, $\tilde{\mathbf{w}}_2\gets \mathbf{0}$}
\ForEach{$\mathbf{w}_1, \mathbf{w}_2 \in \mathbf{W}$}{
  $g_1(\varphi, \theta)\gets\mathbf{w}^H_1\textbf{a}(\varphi, \theta)$\;
  $g_2(\varphi, \theta)\gets\mathbf{w}^H_2\textbf{a}(\varphi, \theta)$\;
  $|g(\varphi, \theta)|\gets \sqrt{\frac{|g_1(\varphi, \theta)|^2+|g_2(\varphi, \theta)|^2}{2}}$\;
  Compute $\sigma^2_g$ in terms of Eq. (\ref{eqn:beamVariation})\;
  \If{$\sigma^2_g<\tilde{\sigma}^2_g$}{
    $\tilde{\sigma}^2_g \gets \sigma^2_g$,
    $\tilde{\mathbf{w}}_1 \gets \mathbf{w}_1$,
    $\tilde{\mathbf{w}}_2 \gets \mathbf{w}_2$\;
  }
  \If{$\tilde{\sigma}^2_g==0$}{
  Stop, \KwRet{$\tilde{\mathbf{w}}_1$, $\tilde{\mathbf{w}}_2$}\;
  }
}
\end{algorithm}

\subsection{Implementation}

There are three types of architecture: digital, analog, and hybrid beamforming. Implementing digital beamforming in a mmWave or THz transceiver equipped with a large-scale array needs a large number of RF components, leading to high hardware cost and power consumption. This constraint has driven the application of analog beamforming, using only a single RF chain. Analog beamforming is implemented as the \textit{de-facto} approach for indoor mmWave systems. However, it only supports single-stream transmission and suffers from the hardware impairment of analog phase shifters. As a trade-off, hybrid beamforming \cite{Ref_zhang2019hybrid} has been proposed as an efficient approach to support multi-stream transmission with only a few RF chains and a phase-shifter network.

The weighting vector $w_{n_xn_y}(t)$ in (\ref{Eqn_UPAbeam}) can be implemented by multiplexing a weighting coefficient over each baseband branch in digital beamforming or adjusting the signal phase on each antenna directly by an analog shifter. Hence, the proposed scheme is applicable for all three kinds of beamforming.
\subsubsection{Digital Beamforming}
Firstly, a physical antenna array is divided into two sub-arrays virtually, e.g., dividing an N-element  array denoted by $\textbf{e}=\{e_1,e_2,\ldots,e_N\}$ as
\begin{equation}
    \begin{cases}
     \textbf{e}_1 =\{e_1,e_2,\ldots,e_{N/2}\}& \\
     \textbf{e}_2=\{e_{N/2+1},e_{N/2+2},\ldots,e_N\}&
\end{cases}
\end{equation} A basis weighting vector  is determined for either sub-array, and this pair of beam patterns complement each other so that their composite pattern is isotropic instantaneously at any particular angle rather than statistically equal by averaging over many TFBs in the RBF.
In conventional beamforming, a data symbol is weighted correctly and transmitted at all elements to achieve constructively superposition in desired directions. In the complementary beamforming, two streams are independently beam formed over $\textbf{e}_1$ and $\textbf{e}_2$, respectively. The electromagnetic interference phenomenon occurs only among elements transmitting correlated signals. Thereby, the pattern of different sub-arrays can be regarded as independent.
\subsubsection{Hybrid Beamforming}
There are at least two RF chains in hybrid architecture. If the number of RF chains is even, we can group them into pairs, and apply the CBF directly  pair-by-pair. Two  physical sub-arrays corresponding to each pair of RF chains generates a pair of complementary beams. If the number of RF chains is odd, we can group the last three RF chains into one group, and then find three complementary beams to minimize the deviation of their composite pattern.
\subsubsection{Analog Beamforming} It can generate a single pattern at each TFB since it has only one RF chain. We can apply a pair of complementary patterns over two consecutive TFBs, similar to the RBF. The major difference is that two consecutive patterns in the RBF is random, rather than complementary.


\begin{figure*}[!tbph]
\centerline{\hspace{0mm}
\subfloat[]{\includegraphics[width=0.37\textwidth]{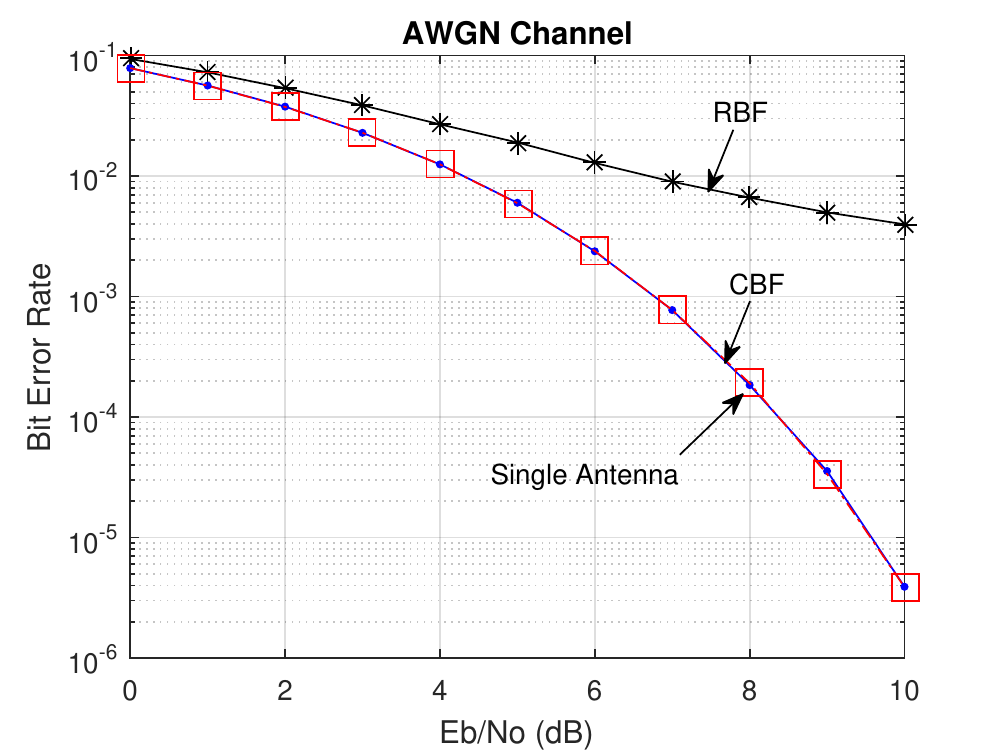}\label{fig:result1}} \\
\subfloat[]{\includegraphics[width=0.37\textwidth]{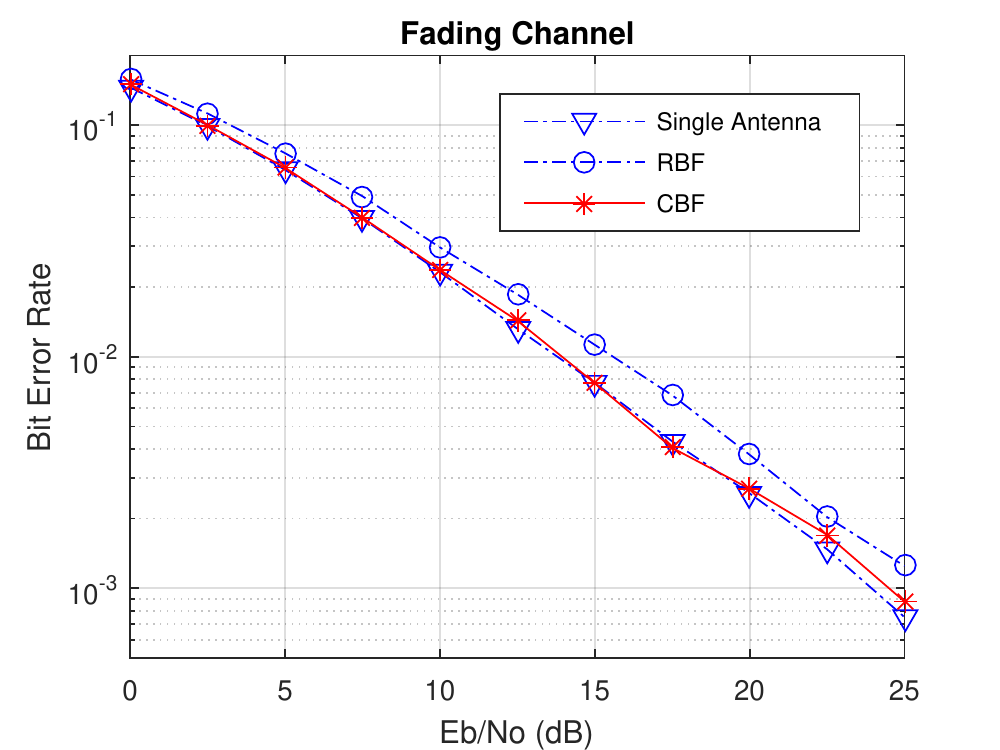}\label{fig:result2}} }
\hspace{0mm}
 \caption{Comparison of uncoded BER performance among the CBF, RBF, and single-antenna broadcasting in (a) Additive White  Gaussian Noise (AWGN) channels and (b) Frequency-flat Rayleigh fading channels.  }
\label{Diagram_Results}
\end{figure*}

\section{Simulations}
The performance of the proposed CBF in terms of the bit error rate (BER) is verified through numerical results acquired by computer simulations using the parameters as summarized in Table \ref{table_SimParameters}. Since link reliability is more important than the data rate in broadcasting synchronization signals and system information, low-order modulation, i.e., quadrature phase-shift-keying (QPSK), is employed. To get an insight into the performance without any other influential factors, the uncoded BER comparison in Additive White Gaussian Noise (AWGN) channels is conducted. Moreover, the results over Rayleigh fading channels are observed, where channel coefficients vary in block-wise.  For a direct comparison with the RBF scheme, an 8-element ULA is applied.

\begin{table}[!t]
\renewcommand{\arraystretch}{1.3}
\caption{Simulation Parameters}
\label{table_SimParameters}
\centering
\begin{tabular}{c|c}
\hline
\textbf{Parameters} & \textbf{Values}  \\
\hline \hline
Antenna array  &  Eight-element ULA, $d=\lambda/2$ \\ \hline
Modulation  &  QPSK  \\ \hline
Channel  & AWGN and flat-fading Rayleigh \\ \hline
Channel estimation & Perfect channel coefficients \\ \hline
Block length & 100 symbols \\ \hline
Simulation length & $10^7$ bits \\ \hline
\end{tabular}
\end{table}

The proposed scheme is compared with the RBF, using the signal-antenna broadcasting as the benchmark.  Note that the single antenna needs a high power amplifier, which brings high hardware costs and power consumption. Moreover, mmWave and THz communications must rely on beamforming over a large-scale array to extend the communication range for transmitting both user data and control signalling. Although a single-antenna broadcasting scheme is optimal for synchronization and broadcast signals, it is not an option for user data transmission in mmWave and THz communications. Hence, it can only serve as the benchmark for omni-directional coverage, rather than a real candidate. In order to verify the omnidirectional coverage property of the CBF, we select different angles for a mobile station to observe its BER performance in the downlink. As we expect for the broadcast channels, the proposed scheme gets identical performance in any angle, as the RBF and the single-antenna broadcasting. However, the CBF outperforms the RBF since it provides instantaneous equal gains in all directions and avoids the fluctuation of energy distribution in the angular domain.

The total power budget for the three schemes is the same for a fair comparison. The uncoded BER performance in an AWGN and frequency-flat Rayleigh fading channel is illustrated in \figurename \ref{Diagram_Results}. The CBF can achieve the optimal performance indicated by the single antenna in both channels since it provides instantaneous equal gain without any beam null. The bit errors of the RBF in AWGN is raised from the beam nulls that cause a very low signal-to-noise ratio. In fading channels, all three schemes suffer from channel fading, where most of the bit errors are caused by the deep fade of wireless channels, but the CBF is also superior to the RBF.

\section{Conclusion}
This paper proposed a novel beamforming technique for the initial access in 6G millimeter-wave and terahertz communications. It provides omni-directional coverage for the broadcasting of synchronization signals and master system information, such as Synchronization Signal Block specified in 5G New Radio, to all mobile uses at any angle of a cell. Unlike the previous random-beamforming technique that achieves omni-directional coverage by averaging many random patterns, it can generate instantaneously isotropic radiation to avoid the energy fluctuation in the angular domain. Consequently, it outperforms random beamforming since the performance loss is caused by the low signal-to-noise ratio due to beam nulls. This scheme can be implemented in digital, analog, and hybrid architecture, and is general for all forms of antenna arrays.

\bibliographystyle{IEEEtran}
\bibliography{IEEEabrv,Ref_WCNC2022}

\begin{thebibliography}{10}
\providecommand{\url}[1]{#1}
\csname url@samestyle\endcsname
\providecommand{\newblock}{\relax}
\providecommand{\bibinfo}[2]{#2}
\providecommand{\BIBentrySTDinterwordspacing}{\spaceskip=0pt\relax}
\providecommand{\BIBentryALTinterwordstretchfactor}{4}
\providecommand{\BIBentryALTinterwordspacing}{\spaceskip=\fontdimen2\font plus
\BIBentryALTinterwordstretchfactor\fontdimen3\font minus
  \fontdimen4\font\relax}
\providecommand{\BIBforeignlanguage}[2]{{%
\expandafter\ifx\csname l@#1\endcsname\relax
\typeout{** WARNING: IEEEtran.bst: No hyphenation pattern has been}%
\typeout{** loaded for the language `#1'. Using the pattern for}%
\typeout{** the default language instead.}%
\else
\language=\csname l@#1\endcsname
\fi
#2}}
\providecommand{\BIBdecl}{\relax}
\BIBdecl

\bibitem{Ref_jiang2022kickoff}
W.~Jiang and H.~D. Schotten, ``The kick-off of {6G} research worldwide: An
  overview,'' in \emph{Proc. 2021 Seventh IEEE Int. Conf. on Comput. and
  Commun. (ICCC)}, Chengdu, China, Dec. 2021.

\bibitem{Ref_jiang2017experimental}
W.~Jiang \emph{et~al.}, ``Experimental results for {Artificial
  Intelligence}-based self-organized {5G} networks,'' in \emph{Proc. {IEEE}
  Int. Symp. on Pers., Indoor and Mobile Radio Commun. (PIMRC)}, Montreal,
  Canada, Oct. 2017.

\bibitem{Ref_jiang2021road}
------, ``The road towards {6G}: A comprehensive survey,'' \emph{IEEE Open J.
  Commun. Society}, vol.~2, pp. 334--366, Feb. 2021.

\bibitem{Ref_kuerner2020impact}
T.~Kuerner and A.~Hirata, ``On the impact of the results of {WRC 2019 on THz}
  communications,'' in \emph{Proc. 2020 Third Int. Workshop on Mobile Terahertz
  Syst. (IWMTS)}, Essen, Germany, Jul. 2020.

\bibitem{Ref_siles2015atmospheric}
G.~A. Siles, J.~M. Riera, and P.~G. del Pino, ``Atmospheric attenuation in
  wireless communication systems at millimeter and {THz} frequencies,''
  \emph{{IEEE} Antennas Propag. Mag.}, vol.~57, no.~1, pp. 48 -- 61, Feb. 2015.

\bibitem{Ref_giordani2016initial}
M.~Giordani, M.~Mezzavilla, and M.~Zorzi, ``Initial access in {5G mmWave}
  cellular networks,'' \emph{{IEEE} Commun. Mag.}, vol.~54, pp. 40 -- 47, Nov.
  2016.

\bibitem{Ref_barati2016initial}
C.~N. Barati \emph{et~al.}, ``Initial access in millimeter wave cellular
  systems,'' \emph{{IEEE} Trans. Wireless Commun.}, vol.~15, no.~12, pp. 7926
  -- 7940, Sep. 2016.

\bibitem{Ref_yang2013random}
X.~Yang, W.~Jiang, and B.~Vucetic, ``A random beamforming technique for
  omnidirectional coverage in multiple-antenna systems,'' \emph{{IEEE} Trans.
  Veh. Technol.}, vol.~62, no.~3, pp. 1420 -- 1425, Mar. 2013.

\bibitem{Ref_yang2012methodUS8170132}
X.~Yang and W.~Jiang, ``Method and apparatus for transmitting signals in a
  multiple antennas system,'' U.S. Patent 8\,170\,132, May 1, 2012.

\bibitem{Ref_jiang2012methodUS13685426}
W.~Jiang and X.~Yang, ``Method and apparatus for transmitting broadcast
  signal,'' U.S. Patent Application 13/685\,426, Nov. 26, 2012.

\bibitem{Ref_yang2013methodUS8537785}
X.~Yang and W.~Jiang, ``Method and apparatus for cell/sector coverage of a
  public channel through multiple antennas,'' U.S. Patent 8\,537\,785, Sep. 17,
  2013.

\bibitem{Ref_jiang2012enhanced}
W.~Jiang and X.~Yang, ``An enhanced random beamforming scheme for signal
  broadcasting in multi-antenna systems,'' in \emph{Proc. 2012 IEEE 23rd Int.
  Symp. on Pers., Indoor and Mobile Radio Commun. (PIMRC)}, Sydney, NSW,
  Australia, Sep. 2012, pp. 2055 -- 2060.

\bibitem{Ref_yang2012methodUS13654743}
X.~Yang and W.~Jiang, ``Method, apparatus, and system for controlling
  multi-antenna signal transmission,'' U.S. Patent Application 13/654\,743,
  Oct. 18, 2012.

\bibitem{Ref_jiang2012suppressing}
W.~Jiang and M.~Schellmann, ``Suppressing the out-of-band power radiation in
  multi-carrier systems: A comparative study,'' in \emph{Proc. 2012 IEEE Global
  Commun. Conf. (GLOBECOM)}, Anaheim, CA, USA, Dec. 2012, pp. 1477 -- 1482.

\bibitem{Ref_dahlman20215gNR}
E.~Dahlman, S.~Parkvall, and J.~Sköld, \emph{{5G NR - The Next Generation
  Wireless Access Technology}}.\hskip 1em plus 0.5em minus 0.4em\relax London,
  the United Kindom: Academic Press, Elsevier, 2021.

\bibitem{Ref_chen2013when}
J.~Chen, ``When does asymptotic orthogonality exist for very large arrays?'' in
  \emph{Proc. 2013 IEEE Global Commun. Conf. (GLOBECOM)}, Atlanta, USA, Dec.
  2013, pp. 4146--4150.

\bibitem{Ref_yang2011random}
X.~Yang, W.~Jiang, and B.~Vucetic, ``A random beamforming technique for
  broadcast channels in multiple antenna systems,'' in \emph{Proc. 2011 IEEE
  Veh. Techno. Conf. (VTC Fall)}, San Francisco, USA, Sep. 2011.

\bibitem{Ref_zhang2019hybrid}
J.~Zhang \emph{et~al.}, ``Hybrid beamforming for {5G} and beyond
  millimeter-wave systems: A holistic view,'' \emph{IEEE Open J. Commun.
  Society}, vol.~1, pp. 77 -- 91, 12 2019.

\end{thebibliography}

\end{document}